\documentstyle[12pt]{article}
\begin{document}
\title{Simple denoising algorithm using wavelet transform}
\footnotetext[1]{e-mail address for correspondence: ravi@che.ncl.res.in}
\author{Manojit Roy,
V. Ravi Kumar$^1$, B. D. Kulkarni \\
{\it Chemical Engineering Division, National Chemical Laboratory,} \\
{\it Pune 411 008, India} \\
John Sanderson, Martin Rhodes \\
{\it Department of Chemical Engineering, Monash University,} \\
{\it Clayton, Victoria, 3168, Australia} \\
Michel vander Stappen \\
{\it Unilever Research, Vlaardingen, Postbox 114, 3130,} \\
{\it AC Vlaardingen, The Netherlands}}
\baselineskip 20pt
\parskip 20pt
\maketitle

\noindent
{\bf keywords:} Noise, Discrete wavelet transform, Chaos, Differentiation

\vspace{5pt}

Application of wavelets and multiresolution analysis to reaction
engineering systems from the point of view of process monitoring, fault
detection, systems analysis etc. is an important topic and of current
research interest (see, Bakshi and Stephanopoulos, 1994; Safavi et.
al., 1997; Luo et. al., 1998; Carrier and Stephanopoulos, 1998). 
In the present paper we focus on one such important
application, where we propose a new and simple algorithm for the
reduction of noise from a scalar time series data.
Presence of noise in a time--varying signal restricts one's ability
to obtain meaningful information from the signal. Measurement of
correlation dimension can get affected by a noise level as small as
1\% of signal, making estimation of invariant
properties of a dynamical system, such as the dimension of the
attractor and Lyapunov exponents, almost impossible (Kostelich and
Yorke, 1988). Noise in experimental data
can also cause misleading conclusions (Grassberger et. al., 1991).
A host of literature exists on various techniques
for noise reduction (Kostelich and Yorke, 1988; H\"ardle, 1990; Farmer
and Sidorowich, 1991; Sauer, 1992; Cawley and Hsu, 1992; Cohen, 1995;
Donoho and Johnstone, 1995; Kantz and Schreiber, 1997). For instance,
Fast Fourier Transform (FFT) reduces noise effectively
in those cases where the frequency distribution of noise is known
(Kostelich and Yorke, 1988; Cohen, 1995; Kantz and Schreiber, 1997); 
singular value analysis methods (Cawley and Hsu, 1992) project the 
original time--series onto an optimal subspace, whereby 
noise components are left behind in the remaining orthogonal
directions, etc. In the 
existing wavelet--based denoising methods (Donoho and Johnstone, 1995) 
two types of
denoising are introduced: linear denoising and nonlinear denoising.
In linear denoising, noise is assumed to be concentrated only on the fine
scales and all the wavelet coefficients below these scales are cut off.
Nonlinear denoising, on the other hand, treats noise reduction by
either cutting off all coefficients below a certain threshold
(so called `hard--thresholding'), or reducing all coefficients by
this threshold (so called `soft--thresholding'). The threshold values
are obtained by statistical calculations and has been seen to depend on
the standard deviation of the noise (Nason, 1994).

The noise reduction algorithm that we propose here makes use of the 
wavelet transform (WT) which in many ways complements the well known
Fourier Transform (FT) procedure. We apply our method, firstly, 
to three model flow systems, {\it viz.} Lorenz, Autocatalator,
and R\"ossler systems, all exhibiting chaotic dynamics.
The reasons for choosing these systems are the following: Firstly, all of 
them are simplified models of well--studied experimental systems. 
For instance, Lorenz is a simple realization of convective
systems (Lorenz, 1963), while the Autocatalator and R\"ossler have 
their more complicated analogs in chemical multicomponent reactions
(R\"ossler, 1976; Lynch, 1992). Secondly, chaotic dynamics is extremely
nonlinear, highly sensitive, possesses only short--time correlations
and is associated with a broad range of frequencies (Guckenheimer and 
Holmes, 1983; Strogatz, 1994). Because of these properties it is well
known that FT methods are not applicable in a straightforward way to
chaotic dynamical systems (Abarbanel, 1993). On the other hand, WT
methods are particularly suited to handle not only nonlinear but
nonstationary signals (Strang and Nguyen, 1996). This is because the
properties of the data are studied at varying scales with superior time
localization analysis when compared to FT technique. Our noise
reduction algorithm is advantageous, because, as shall be shown, the
threshold level for noise is identified automatically. In this study,
we have used the discrete analog of the wavelet transform (DWT) which
involves transforming a given signal with orthogonal wavelet basis
functions by dilating and translating in discrete steps (Daubechies,
1990; Holschneider, 1995). For study purposes we corrupt
one variable $x(t)$ for each of these systems with
noise of zero mean, and then apply our algorithm for denoising. We 
analyze the performance of this method in all the three systems
for a wide range of noise strengths, and show its effectiveness.
Importantly, we then validate the applicability of the method to 
experimental
data obtained from two chemical systems. In one system the time series 
data was obtained from pressure fluctuation measurements of the 
hydrodynamics in a fluidized bed. In the other the conductivity 
measurements in a liquid surfactant manufacturing experiment were 
analyzed.

\noindent
{\bf Methodology}

\noindent
The noise reduction algorithm based on DWT consists of the following
five steps:

\noindent
{\it Step 1}: 

\noindent
In first step, we differentiate the noisy signal $x(t)$ to obtain 
the data $x_d(t)$, using the method of central finite 
differences with fourth order correction to minimize error 
(Constantinides, 1987), {\it i.e.}, 
\begin{eqnarray}
x_d(t) \;\;\;\; = \;\;\;\; \frac{d x(t)} {d t}\;. \label{e1}
\end{eqnarray}

\noindent
{\it Step 2}:

\noindent
We then take DWT of the data $x_d(t)$ and obtain wavelet coefficients 
$W_{j,k}$ at various {\it dyadic scales} $j$ and displacements
$k$. A dyadic scale is the scale whose numerical magnitude is equal to
2 (two) raised to an integer exponent, and is labeled by the exponent.
Thus, the dyadic scale $j$ refers to a scale of magnitude $2^j$. In
other words, it indicates a resolution of $2^j$ data points. Thus a low
value of $j$ implies finer resolution while high $j$ analyzes the
signal at coarser resolution.
This transform is the discrete analog of continuous WT 
(Holschneider, 1995), and is given by the formula
\begin{eqnarray}
W_{j,k} \;\;\; = \;\;\; \int_{-\infty}^{+\infty} x_d(t)
		 \; \psi_{j,k} (t)\;d t\;, \label{e2}
\end{eqnarray} 
with 

\centerline{$ \psi_{j,k}(t) \;\; = \;\; 2^{j/2} \psi(2^jt - k) $}
\noindent
where $j,k$ are integers. As for
the wavelet function $\psi(t)$ we have chosen Daubechies compactly
supported orthogonal
function with four filter coefficients (Daubechies, 1990; 
Press et. al., 1996). 

\noindent
{\it Step 3:}

\noindent
In this step we estimate the {\it power} $P_j$
contained in different dyadic scales $j$, via
\begin{eqnarray}
P_j(x) \;\;\;\; = \;\;\;\; \sum_{k=-\infty}^{+\infty} 
	\vert W_{j,k} \vert^2 \;\;\;\;\; (j = 1, 2, \cdots) \label{e3}
\end{eqnarray}
By plotting the variation of $P_j$
with $j$, we see that it is possible to identify a scale $j_m$ at which
the power due to noise falls off rapidly. This is important because as
we shall see from the studies of the case examples that it provides a
mean for automation in detection of threshold. The identification of
the scale $j_m$ at which power due to noise shows the first minimum
allows us to reset all $W_{j,k}$ upto scale $j_m$ to zero, {\it i.e.},
$W_{j,k} = 0,$ for $j = 1, 2, \cdots, j_m$. 

\newpage

\noindent
{\it Step 4:}

\noindent
In the fourth step, 
we reconstruct the denoised data $\hat x_d(t)$ by taking inverse 
transform of the coefficients $W_{j,k}$ : 
\begin{eqnarray}
\hat x_d(t) \;\;\;\; = \;\;\;\; c_\psi \sum_{j=0}^{\infty}
	 \sum_{k=-\infty}^{\infty} W_{j,k} \; \psi_{j,k}(t)\;, \label{e4} 
\end{eqnarray}
where $c_\psi$ is normalization constant given by

\centerline{$ c_\psi \;\; = \;\; 1 / \int_{-\infty}^{\infty} \frac{ \vert
	\hat \psi(\omega) \vert^2 }{ \omega} d \omega \; < \; \infty\;, $}
\noindent
with $\hat \psi(\omega)$ as the Fourier transform of the wavelet 
function $\psi(t)$.

\noindent
{\it Step 5:}

\noindent
In the fifth and final step $\hat x_d(t)$
is integrated to yield the cleansed signal $\hat x(t)$:
\begin{eqnarray}
\hat x(t) \;\;\;\; = \;\;\;\; \int \hat x_d(t) d t \; . \label{e5}
\end{eqnarray}

There exists a commutativity property between the operation of
differentiation/integration and
wavelet transform. Therefore first differentiating the signal
and then taking DWT is equivalent to carrying out
the two operations in reverse order. This implies that the same
result can be obtained by switching the order between the first and 
second steps, and then between the fourth and fifth.

The effectiveness of the method lies in the following observations.
Upon differentiation, contribution due to white noise moves towards the
finer scales because the process of differentiation converts the
uncorrelated stochastic process to a first order moving average process
and thereby distributes more energy to the finer scales. That the
differentiation of white noise brings about this behavior is known in
the Fourier spectrum (Box et. al., 1994). It may be noted that the
nature and effectiveness of separation depend on the wavelet basis
function chosen and also on the properties of the derivatives of WT, which
is in itself a highly interesting and not fully understood subject
(Strang and Nguyen, 1996). For the signals studied in this paper, model
as well as experimental, the wanted signal features lie in the coarser
wavelet scales while the unwanted signal features after differentiation
lie in the finer resolution wavelet scales. This is because the size of
the data set handled decides the total number of scales available and a
suitable choice can bring out the noisy signal WT features lying in the
coarser scales. This justifies the assumption that fine scale features
can be removed by setting the corresponding wavelet coefficients to
zero and coarse scale features retained after differentiation. For this
reason we also see a clear separation in the scales attributed to
noise and those for the signal. A threshold scale for noise removal is
thus identified and this leads to an automation for noise removal. 

\noindent
{\bf Result and Discussion}

\noindent
We first take up the three
model systems and discuss the observations. For test purposes the
pure signal obtained from these systems were corrupted with noise of
certain strength. For the systems chosen for study, {\it viz.}, Lorenz, 
Autocatalator, and R\"ossler, Table~\ref{t1} summarizes the details,
{\it i.e.} the equations governing their dynamics, the values 
chosen for the parameters, and the nature of the evolution of these
systems for these sets of parameter values. These values were chosen
appropriately so that the dynamics is chaotic. In our initial studies,
purely for testing purposes, we studied situations where we ensure that
all scales are affected by noise. In the wavelet domain this can be
conveniently carried out by perturbing the wavelet coefficients in the
following way. The differential equations
are first numerically integrated to obtain pure signal
$x^0(t_i)$ at equidistant time steps $t_i$. We then take DWT of the 
signal, and add white noise $\eta$ of zero mean
and certain strength, {\it i.e.}, $W_{j,k} = W_{j,k}^0 + \eta$, where
$W_{j,k}^0, W_{j,k}$ are the wavelet coefficients of pure and noisy
signals respectively. 
We take the strength of noise as the relative percentage of the
difference between the maximum and minimum of the signal value.
Since each coefficient $W_{j,k}$ is individually affected by the noise,
this procedure ensures equal weightage for presence of noise at all scales.
Reconstructing the time series signal with this perturbed set of
wavelet coefficients gave us the noisy signal to be cleansed. Our
studies in this fashion did show that noise and signal separation was
achieved. For the subsequent studies we followed the usual way of
corrupting the signal by additive noise, {\it i.e.},
\begin{eqnarray}
x(t_i) = x^0(t_i) + \eta(t_i)\;, \label{e6}
\end{eqnarray}
where $\eta(t_i) \in [-.5, .5]$ is the noise with zero mean 
and uniform distribution, and $J$ the number of available dyadic 
scales. We have taken data size of 16384 ($= 2^{14}$) points for all 
these three systems, and so $J = 14$.

In Fig.~1 we plot our observations for the Lorenz system.
Fig.~1 (a) shows the power at different scales in the pure signal
$x^0(t_i)$. In Fig.~1 (b) we plot the scalewise power distribution
after numerically differentiating the pure signal. We see that almost
the entire power of the differentiated data is accumulated within 
the dyadic scales 4 and 9 (the signal power between scales 10 and 14
has disappeared by the process of differentiation).
Fig.~1 (c) plots the scalewise power in the noisy signal
$x(t_i)$ when the pure signal is infected with 
noise $\eta(t_i)$ of a typical
strength of 5\% of the signal (that is, 5\% of the difference in the
maximum and minimum values of $x^0(t_i)$). Because of the relative
larger contribution at all scales from the pure signal, compared to
that from noise, it is impossible to distinguish between
the two components, and the figure looks qualitatively very similar to 
Fig.~1 (a). However, when we plot the scalewise power distribution of the
differentiated noisy data in Fig.~1 (d), the signal contribution can
easily be identified and also compared with the plot in Fig.~1 (b). It is
to be noted that the 
difference in the values of the two peaks in Figs.~1 (b) and (d)
arises because of the power being
normalized by the respective total signal power.
The contribution due to noise shows up in the finer scales.
A clear minimum with close to zero value separates out two distinct
regions. Fig.~1 (e) exhibits a
small segment of the signal after the noise has been
successfully removed following the procedure outlined above.
All the three signals -- pure, noisy, and cleansed
-- are overlaid for the sake of comparison.

In Fig.~2 we show the results for the Autocatalator and the R\"ossler
reacting systems. Fig.~2 (a) and (c) plot, respectively, the scalewise
power 
distribution for noise--infected signals obtained from the two systems. 
Like in the case of Lorenz system, it is evident that here also one
cannot distinguish the noise and signal components.
Fig.~2 (b) and (d) exhibit scalewise power profile for the differentiated 
data of the two signals respectively. The clear separation is again
obvious.

In order to quantitatively estimate the efficiency of our denoising
method, we have made the following error estimation (Kostelich and
Schreiber, 1993)
for the above three model systems. Since in all these cases 
the pure signal is known, a measure of the amount of error present in the
cleaned data is obtained by taking rms deviation of the cleaned
signal $\hat x(t_i)$ from the pure signal $x^0(t_i)$ as follows,
\begin{eqnarray}
\hat E = \Big ( \frac{1}{N} \sum_{i=1}^N
         (\hat x(t_i) - x^0(t_i))^2 \Big )^{1/2}\;, \label{e7}
\end{eqnarray}
where $N$ is the length of the time series.
Similar quantity $E$ for the noisy data $x(t_i)$ is also computed. The
condition $\hat E / E < 1$ guarantees that noise has been successfully
reduced. The error estimator $\hat E / E $ is a natural measure for
noise reduction when the original dynamics is known (Kostelich and
Schreiber, 1993). In Fig.~3 we plot $\hat E / E$ against noise
strength, for the
three model systems. We see that for the entire range of noise
values, and even with noise level as high as 10\% of the signal
exhibiting chaotic dynamics,
$\hat E / E$ remains appreciably below unity. Thus the
plot demonstrates the efficiency of the approach. Different wavelet
basis functions may change the nature and also improve the efficiency
further.

We now discuss our method when applied to raw data obtained from 
two real chemical systems. In the
first system, the time series data was obtained from the measurements 
of the pressure fluctuations in a fluidized bed, which consists of a 
vertical chamber inside of which a bed of solid particles is
supported by an upwardly moving gas. Our system used a bed of silica
sand particles (of mean diameter 200 microns) with a settled height of
500 mm, fluidized by ambient air in a transparent vessel 430 mm across
and 15 mm wide. Beyond a critical inlet gas velocity, {\it viz.} the
minimum bubbling velocity, the gas passes through the bed in the form of 
bubbles, thereby churning the solid and gas mixture in a turbulent
manner. The time series data have been taken by measuring the
pressure fluctuations inside this mixture, relative to atmospheric
pressure, using a pressure transducer attached to a probe inserted into
the fluid bed. The bed was operated at an inlet gas velocity of 0.85
m/sec, and the pressure fluctuations were recorded at a sampling rate
of 333 Hz (333 data points per second). As a standard procedure, we 
normalize the data by
subtracting mean and dividing by standard deviation (Constantinides,
1987; Bai et. al., 1997). In Fig.~4 we show
the results obtained after the data have been subjected to denoising. 
Fig.~4 (a) shows the power distribution at different scales
in the original experimental signal, while in Fig.~4 (b) we plot the 
scalewise
power profile of the differentiated data. Again one clearly sees the two
distinct contributions due to the noise and signal components. Fig.~4 (c)
shows short segments of the denoised signal and the original 
signal which is overlaid for comparison.
The cleaned signal is seen to be smooth indicating that the
noise has been removed.

In the second chemical system, the time series data was obtained by 
sampling a measure related to the conductivity in a 3 liter liquid 
surfactant manufacturing
experiment, at a sampling rate of 500 Hz. The time series is highly 
nonstationary since at various
stages the operational parameters are altered (increasing the
temperature for certain 
duration, then adding actives to the liquid, etc.). We studied
unfiltered noisy data sets from the experiments, in order to check if
our method 
can filter the noise out and also bring forth some intrinsic features of
the system. We used our denoising algorithm to treat this data set
in a slightly different way. The aim was to remove the finer scales
from the differentiated data one
by one, starting from the lowest (dyadic scale 1) and gradually going
up, so that at each stage (after integrating the data) the
observable frequencies in the filtered 
signal may be related to identifiable physical sources. Fig.~5 (a) 
shows a small 
segment (1 second long) of the noisy data. In Fig.~5 (b) we plot, on
the same scale as in the earlier figure, the
filtered data, using our method to remove the lowest dyadic scale 1.
One can now clearly identify a 50 Hz component, due to the signal from
electrical power supply (the `net frequency'). By removing scale 2
alongwith scale 1, the net frequency goes away, and the filtered data
exhibits a 13 Hz component superimposed with occasional spikes. This
13 Hz signal shows up clearly in the filtered data with scale 3 also
removed. The same Fig.~5 (b) shows this data, overlaid on the data with
scale 1 removed. This 13 Hz may have arisen from the stirring device which
has two blades and revolves with 260 rpm, coresponding to approximately
10 Hz. The electronic signal had an antialiasing feature of no more
than 250 Hz and therefore aliasing (beating) may be ruled out.
It may also be mentioned here that the Fourier power 
spectrum of the denoised signal shows a spike at 50 Hz frequency, whose
removal 
resulted in a residual spectrum consisting mainly of a background
continuum without any appreciable peak around 13 Hz. This study with
the present example suggests that the wavelet transform methodology
offers considerable benefits in the recovery of intrinsic signal components.

\noindent
{\bf Summary}

\noindent
We have presented a new and alternative algorithm for noise reduction
using discrete wavelet transform. We believe that our algorithm will be
beneficial in various noise reduction applications, and that it shows
promise in developing techniques which can resolve an observed signal
into its various intrinsic components.
In our method the threshold for reducing
noise comes out automatically. The algorithm has been applied to three
model flow systems - Lorenz, Autocatalator, and R\"ossler systems -
all evolving chaotically. The method is seen to work quite well for
a wide range of noise strengths, even as large as 10\% of the signal
level. We have also applied the method successfully to noisy time
series data
obtained from the measurement of pressure fluctuations in a fluidized
bed, and also to that obtained by conductivity measurement
in a liquid surfactant experiment. In all the illustrations we have
been able to observe that there is a clean separation in the
frequencies covered by the differentiated signal and white noise.
However, if the noise is colored, a certain degree of overlap between
the signal and noise may exist even after differentiation. For this
complex situation, the method needs to be improved upon.

\noindent
{\bf Acknowledgement}

\noindent
Authors acknowledge Unilever Research, Port Sunlight, for financial and
other assistance. Part of the work has been carried out under the aegis
of Indo--Australian S\&T program DST/INT/AUS/I--94/97.

\noindent
{\bf Literature cited}

\noindent
Abarbanel, H. D. I., ``The Observance of Chaotic Data in Physical
Syatems", Rev. Mod. Phys. 65, 1340 (1993)

\noindent
Bai, D., T. Bi, and J. R. Grace ``Chaotic Behavior of Fluidized Beds
Based on Pressure and Voidage Fluctuations", AIChE Journal 43, 1357 (1997)

\noindent
Bakshi, B., and G. Stephanopoulos ``Representation of Process Trends.
Part IV. Induction of Real--time Patterns from Operating Data for
Diagnoses and Supervisory Control", Computers chem. Engng. 18, 303 (1994)

\noindent
Box, G. E. P., G. M. Jenkins and G. C. Reinsel, ``Time Series Analysis
Forecasting and Control" Prentice Hall, Englewood Cliffs (1994)

\noindent
Carrier, J. F., and G. Stephanopoulos ``Wavelet--Based Modulation in
Control--Relevant Process Identification", AIChE Journal 44, 341 (1998)

\noindent
Cawley, R., and G.--H. Hsu ``Local--Geometric--Projection Method for
Noise Reduction in Chaotic Maps and Flows", Phys. Rev. A 46, 3057 (1992)

\noindent
Cohen, L., ``Time frequency analysis" Prentice Hall, Englewood Cliffs (1995)

\noindent
Constantinides, A., ``Applied Numerical Methods with
Personal Computers" McGraw--Hill Book Company, USA (1987)

\noindent
Daubechies, I., ``Ten Lectures on Wavelets" SIAM, Philadelphia (1990)

\noindent
Donoho, D. L., and I. M. Johnstone ``Adapting to Unknown Smoothness via
Wavelet Shrinkage", J. American Statistical Association 90, 1200 (1995)

\noindent
Farmer, J. D., and J. J. Sidorowich ``Optimal Shadowing and Noise
Reduction", Physica D 47, 373 (1991)

\noindent
Grassberger, P., T. Schreiber, and C. Schaffrath ``Non--linear Time 
Sequence Analysis", Int. J. Bifurcation Chaos 1, 521 (1991)

\noindent
Guckenheimer, E. and P. Holmes, ``Nonlinear Oscillation, Dynamical
Systems and Bifurcations of Vector Fields" Springer Verlag, Berlin (1983)

\noindent
H\"ardle, W., ``Applied Nonparametric Regression"
Econometric Society Monographs, Cambridge University Press (1990)

\noindent
Holschneider, M., ``Wavelets: An Analysis Tool" Clarendon Press,
Oxford (1995)

\noindent
Kantz, H. and T. Schreiber, ``Nonlinear Time Series
Analysis" Cambridge University Press, Cambridge (1997)

\noindent
Kostelich, E. J., and J. A. Yorke ``Noise Reduction in Dynamical
Systems", Phys. Rev. A 38, 1649 (1988)

\noindent
Lorenz, E. ``Deterministic Nonperiodic Flow", J. Atmos. Sc. 20, 130 (1963)

\noindent
Luo, R., M. Misra, S. J. Qin, R. Barton, and D. M. Himmelblau ``Sensor
Fault Detection via Multiscale Analysis and Nonparametric Statistical
Inference", Ind. Eng. Chem. Res. 37, 1024 (1998)

\noindent
Lynch, D. T. ``Chaotic Behavior of Reaction Systems: Mixed Cubic
and Quadratic Autocatalators", Chem. Engg. Sci. 47, 4435 (1992)

\noindent
Nason, G. P., ``Wavelet Regression by Cross--validation"
Dept. of Mathematics, University of Bristol (1994)

\noindent
Press, W. H., B. P. Flannery, S. A. Teukolsky and W. T.
Vetterling, ``Numerical Recipes" Cambridge University Press,
Cambridge (1987)

\noindent
R\"ossler, O. E. ``Chaotic Behavior in Simple Reaction Syatems", 
Z. Naturforsch. 31 a, 259 (1976)

\noindent
Safavi, A. A., J. Chen, and J. A. Romagnoli ``Wavelet--Based Density
Estimation and Application to Process Monitoring", AIChE Journal 43,
1227 (1997)

\noindent
Sauer, T. ``A Noise Reduction Method for Signals from Nonlinear
Systems", Physica D 58, 193 (1992)

\noindent
Strang, G. and T. Nguyen, ``Wavelets and Filter Banks" 
Wellesley--Cambridge Press (1996)

\noindent
Strogatz, S. H., ``Nonlinear Dynamics and Chaos: With Applications to
Physics, Biology, Chemistry and Engineering" Addison--Wesley (1994)

\newpage

\centerline{\bf Figure Captions}

\vspace{.2in}

\begin{itemize}

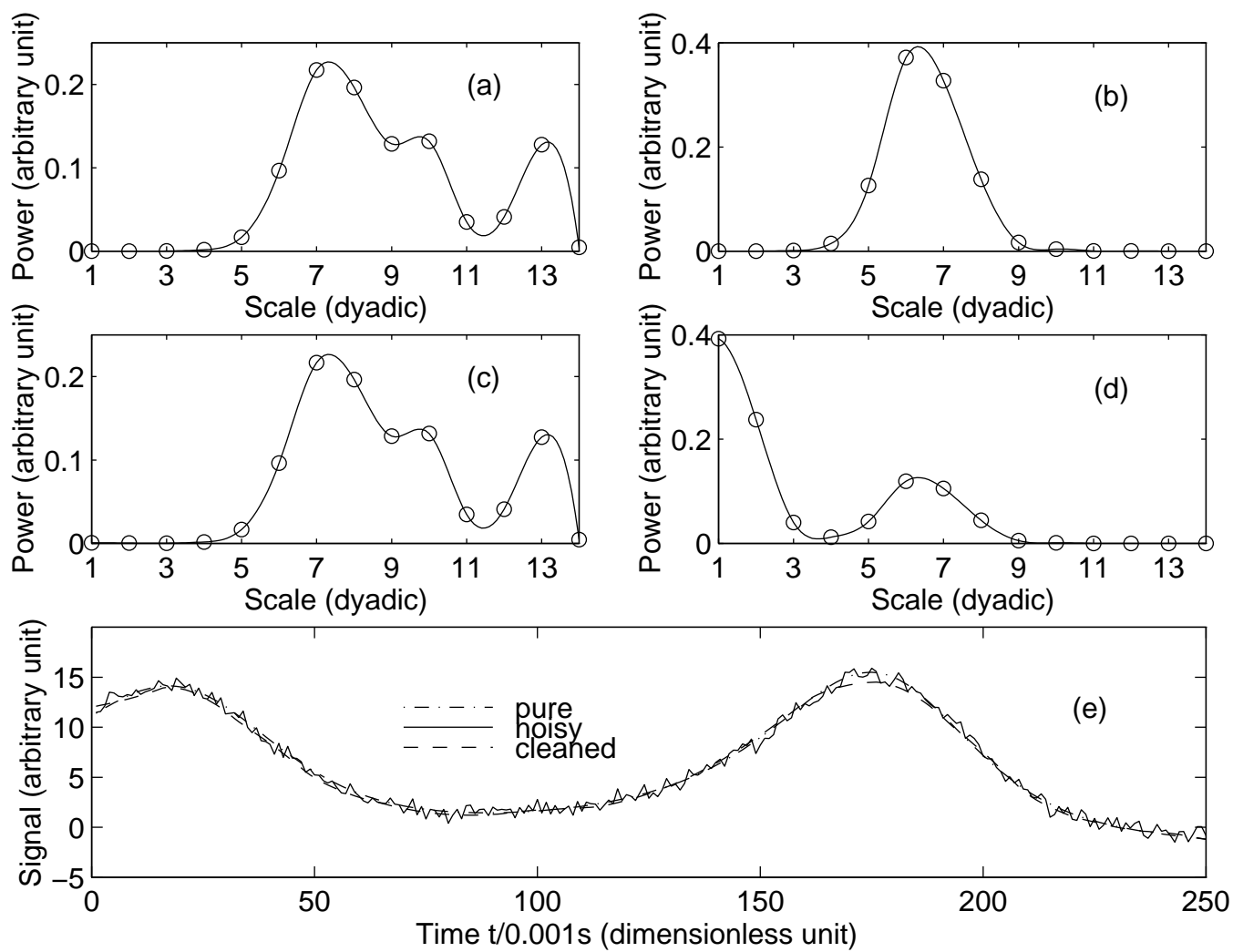
\item[Fig.~1.] Plots for Lorenz system, with parameter values as
     stated in Table~\ref{t1}. 16384
     ($= 2^{14}$) data points are considered. Scalewise power 
     distribution is plotted against the dyadic
     scale, (a) for the pure signal (the interpolated line through the 
     data points is drawn for visualization), (b) using the data 
     obtained after
     differentiating the signal, (c) for the signal corrupted by noise,
     and (d) using differentiated data of noisy signal. A segment of
     cleansed signal is shown in (e), alongwith the pure and noisy
     signals overlaid for comparison.
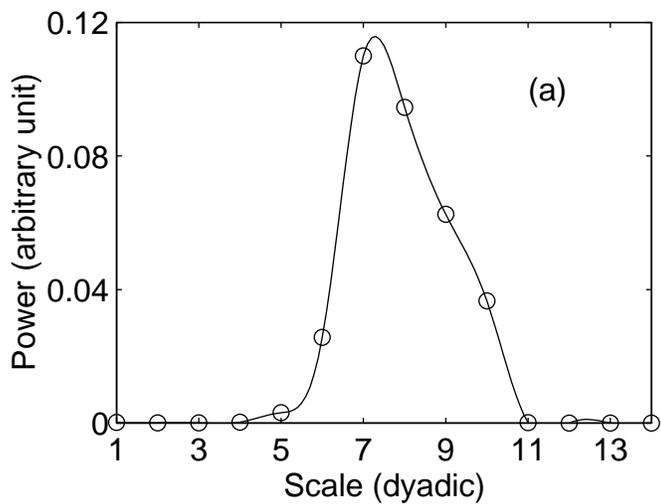
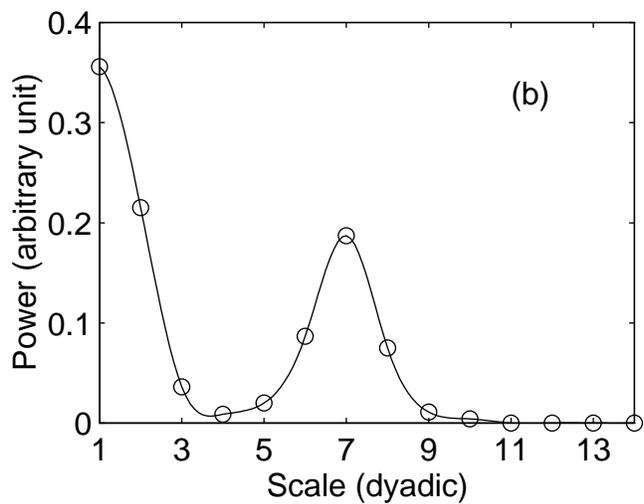
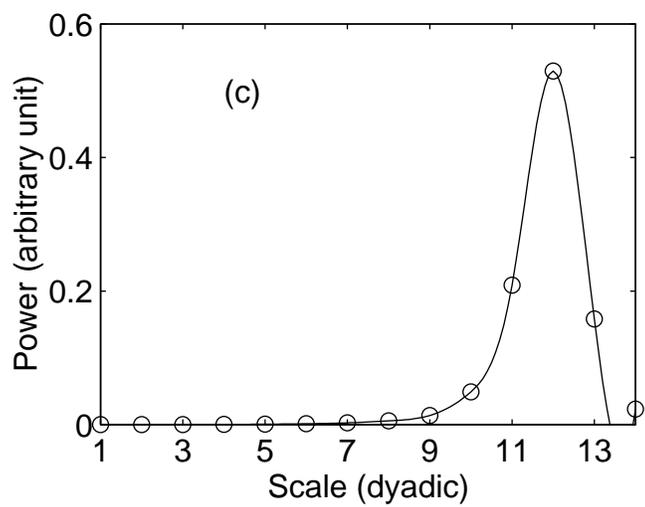
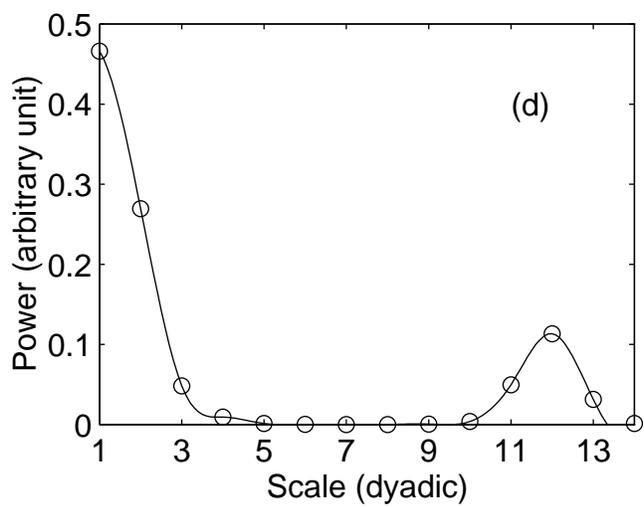
\item[Fig.~2.] Plots for Autocatalator and R\"ossler
     systems, in (a), (b) and (c), (d)
     respectively, for parameter values as in Table~\ref{t1}.
     Scalewise power profile plotted, (a) for the noisy
     autocatalytic signal, (b) using data after differentiating the
     signal, (c) for noisy R\"ossler signal, and (d) using the
     differentiated data of noisy signal.
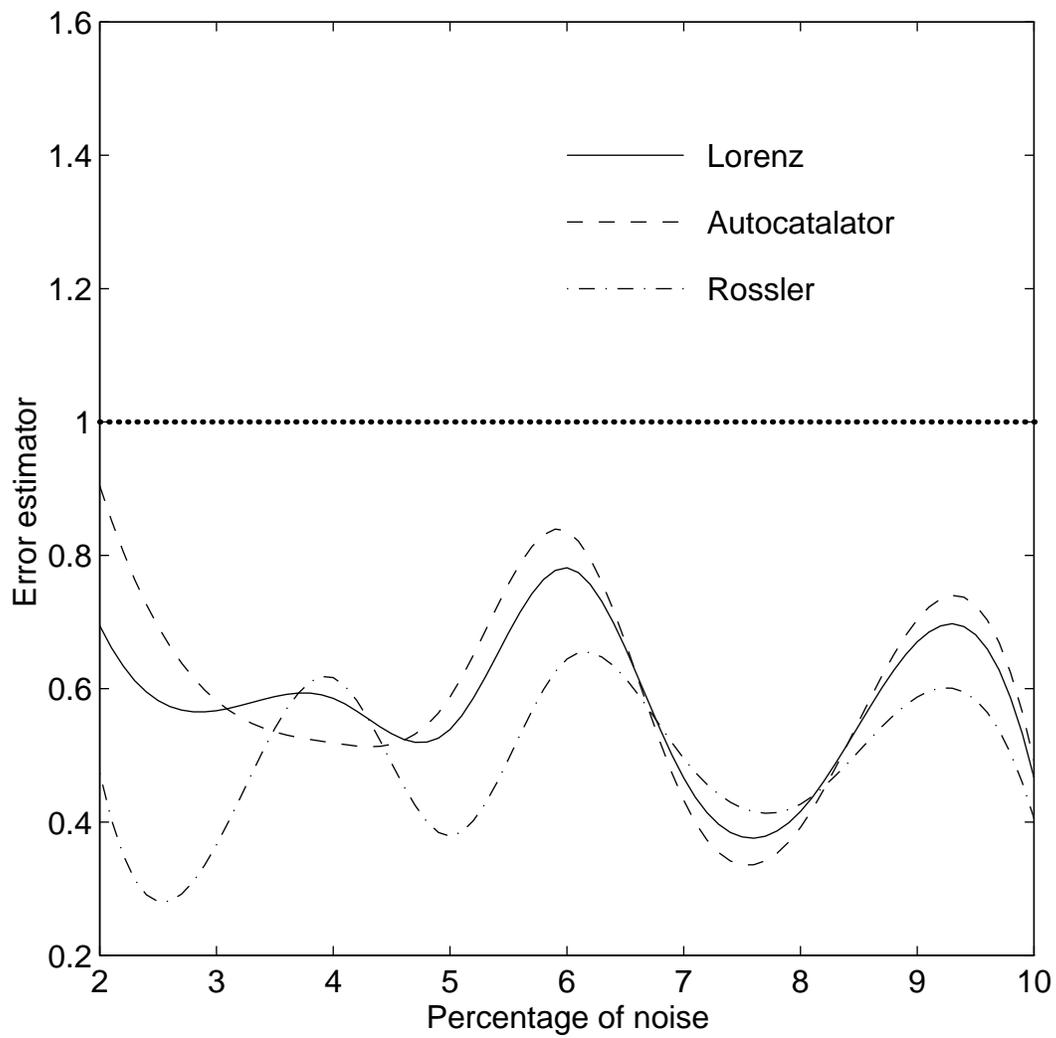
\item[Fig.~3.] The error estimator $\hat E / E$
     plotted against the noise strength for all
     the three systems.
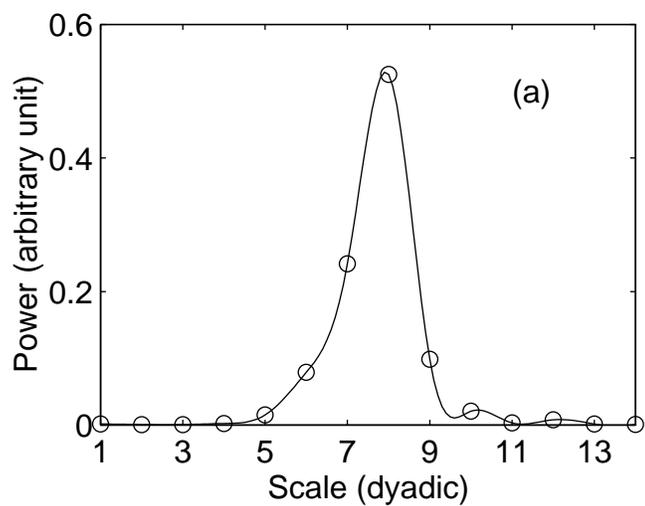
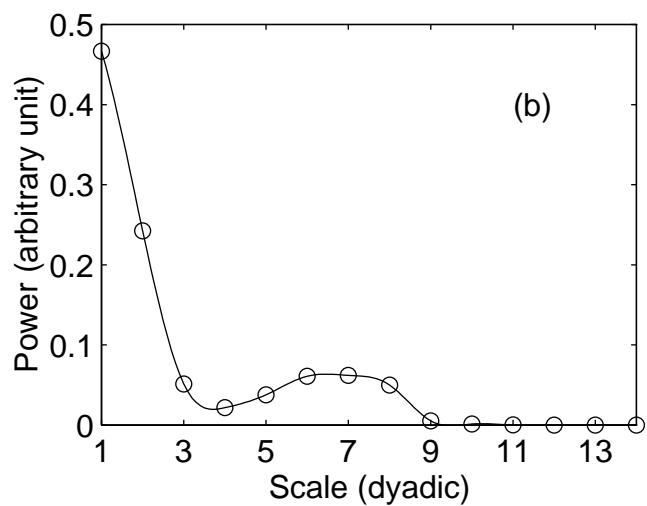
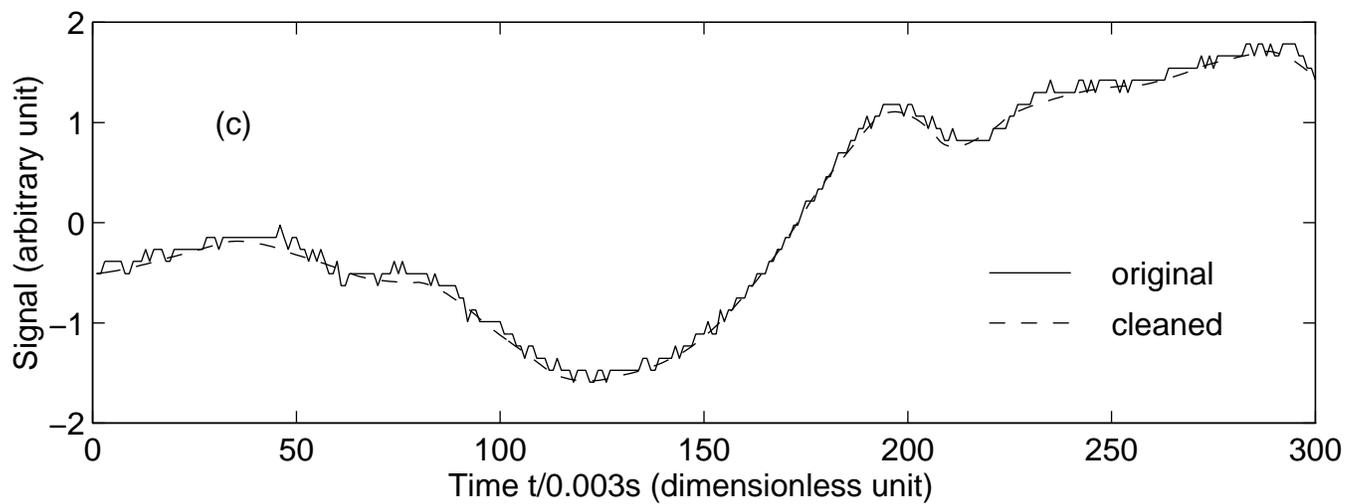
\item[Fig.~4.] Plots for the fluidized bed experiment. (a)
     Scalewise power profile is shown, for (a) the experimental signal,
     and (b) the data after the signal has been numerically
     differentiated. A small segment of the cleansed signal is
     shown in (c), alongwith the original signal for comparison.
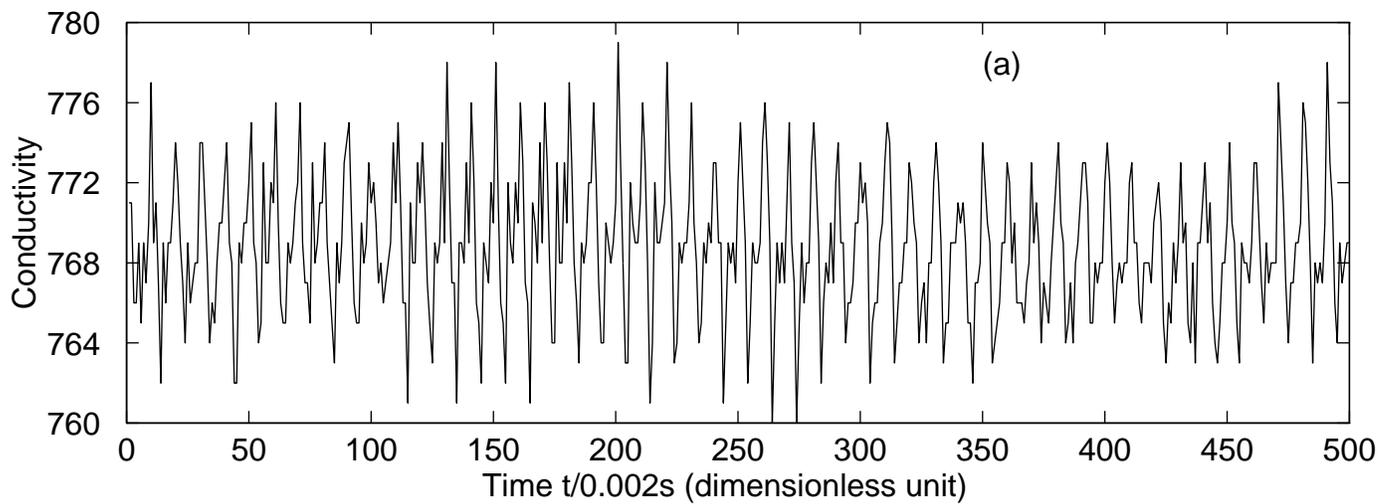
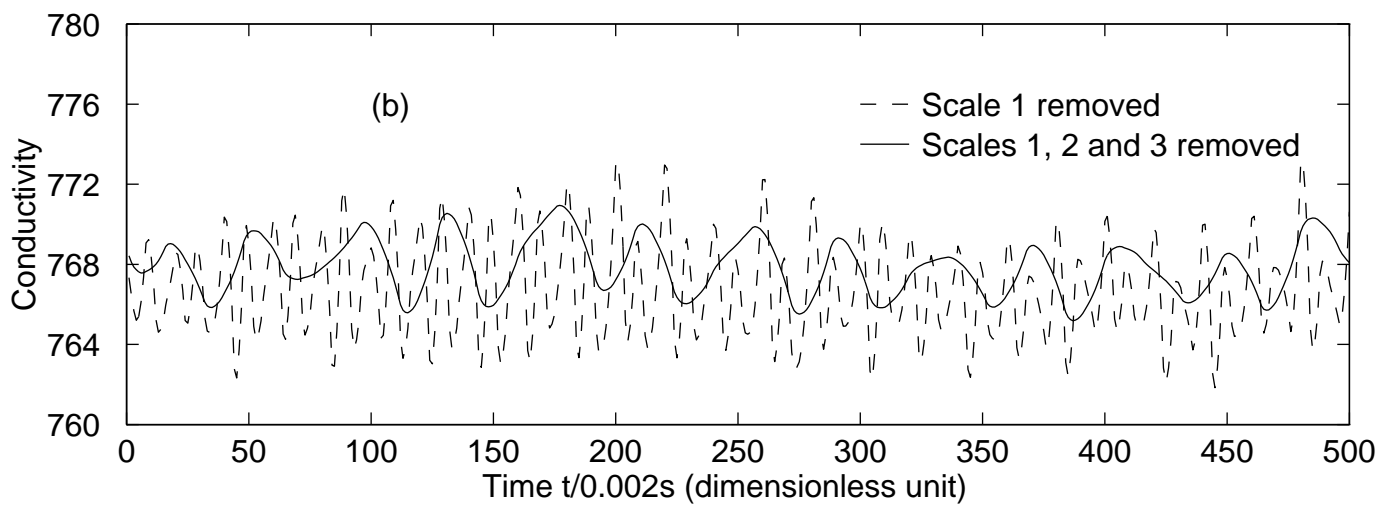
\item[Fig.~5.] Plots for the liquid surfactant experiment. (a) A segment
     of the original noisy data, 1 second long. (b) The filtered data,
     with scale 1 removed, and with scales 1, 2 and
     3 removed (on the same axes--scales as (a)).

\end{itemize}

\newpage

\begin{table}
\caption{Various model systems studied, alongwith their parameter
        values and nature of dynamics.}
\begin{tabular}{||l||l|l|l|} \hline \hline &&&\\
Systems &~~~~~~~~ Lorenz &~~~~~~~ Autocatalator &
			~~~~~ R\"ossler \\ 
&&&\\ \hline
&&&\\
 & $dx/dt = - \sigma x + \sigma y$, 
& $dx/dt = 1 - x - D a_1 x z^2$, & $dx/dt = - y - z$, \\
Dynamical & $dy/dt = R x - y - x z$, 
& $dy/dt = \beta - y - D a_2 y z^2$, & $dy/dt = x + a y$, \\
equations & $dz/dt = - b z + x y$. 
& $dz/dt = 1 - (1 + D a_3) z$ & $dz/dt = b$ \\ 
&&~~~~~ $+ \alpha (D a_1 x + D a_2 y) z^2$. & ~~~~~~ $+ z (x - c)$. \\
&&&\\ \hline
&&&\\
Parameters & $\sigma = 10$, $R = 28$, 
& $D a_1 = 18000$, $D a_2 = 400$, & $a = .398$, $b = 2$, \\
~~~chosen &~~~~ $b = 8/3$. &~~~ $D a_3 = 80$, $\beta = 2.93$, 
&~~~~~ $c = 4$. \\ 
&&~~~~~~~~~~~ $\alpha = 1.5$. &\\
&&&\\ \hline 
&&&\\
Dynamics &~~~~~~~ Chaotic &~~~~~~~~~~ Chaotic &~~~~~ Chaotic \\
&&&\\ \hline \hline
\end{tabular}
\label{t1}
\end{table}

\end{document}